\documentclass[aps,pre,amsmath,amssymb,twocolumn,superscriptaddress,notitlepage]{revtex4-1}
\usepackage{graphicx}
\usepackage[colorlinks = true, linkcolor = black, urlcolor  = blue, citecolor = blue, anchorcolor = blue]{hyperref}
\usepackage{amsmath,amssymb}
\usepackage{flushend}
\usepackage{url}
\usepackage{mathtools}
\usepackage{hyphenat}
\usepackage{color}

\begin{document}
\title{ Anharmonic classical time crystals: A  {coresonance} pattern formation mechanism  }
\author{Zachary G. Nicolaou}
\affiliation{Department of Physics and Astronomy, Northwestern University, Evanston, Illinois 60208, USA}
\author{Adilson E. Motter}
\email[Email: ]{motter@northwestern.edu}
\affiliation{Department of Physics and Astronomy, Northwestern University, Evanston, Illinois 60208, USA}
\affiliation{Northwestern Institute on Complex Systems, Northwestern University, Evanston, Illinois 60208, USA}

\begin{abstract}
Driven many-body systems have been shown to exhibit discrete time crystal phases characterized by broken discrete time-translational symmetry. This has been achieved generally through a subharmonic response, in which the system undergoes one oscillation every other driving period. Here, we demonstrate that classical time crystals do not need to resonate in a subharmonic fashion but instead can also exhibit a continuously tunable anharmonic response to driving, which we show can emerge through a coresonance between modes in different branches of the dispersion relation in a parametrically driven medium. This response, characterized by a typically incommensurate ratio between the resonant frequencies and the driving frequency, is demonstrated by introducing a time crystal model consisting of an array of coupled pendula with alternating lengths. Importantly, the coresonance mechanism is the result of a  bifurcation involving a fixed point and an invariant torus, with no intermediate limit cycles. This bifurcation thus gives rise to {many-body} symmetry breaking phenomenon directly {connecting} the symmetry-unbroken phase with a previously uncharacterized phase of matter, which we call an anharmonic time crystal phase. The mechanism is shown to generalize to driven media with any number of coupled fields and is expected to give rise to anharmonic responses in a range of weakly damped pattern-forming systems, with potential applications to the study of nonequilibrium phases, frequency conversion, and acoustic cloaking.

\vspace{1em}
{\noindent DOI: \href{https://doi.org/10.1103/PhysRevResearch.3.023106}{10.1103/PhysRevResearch.3.023106}}\\
{Phys. Rev. Research \textbf{3}, 023106 (2021)}

\end{abstract}

\maketitle

\section{Introduction}
Inspired by the potential for spontaneous time-translational symmetry breaking \cite{2012_Wilczek}, recent studies have led to the discovery of intriguing nonequilibrium phases of matter known as discrete time crystals \cite{2020_Else}. Continuous time-translational symmetry cannot be broken in closed quantum many-body systems with only local interactions~\cite{2013_Bruno,2015_Watanabe}. However, discrete time crystals can be created in driven systems by exciting period-doubled subharmonic responses to periodic driving, which constitute states of broken discrete time-translational symmetry. Notably, periodically driven chains of quantum spins avoid indefinite heating and exhibit time crystal phases in experimentally realizable systems~\cite{2015_Sacha,2016_Dominic, 2017_Yao, 2017_Zhang}.

While classical time crystals were introduced alongside their quantum counterparts~\cite{2012_Wilczek-2}, they have only more recently garnered significant attention \cite{2018_Goldstein,2018_Yao_Zaletel,2019_Heugel,2019_Gambetta,2020_Hurtado,2020_Libal}.  One of the best characterized classical systems that breaks discrete time-translational symmetry is a swinging pendulum driven by the sinusoidal motion of its support, as described by the Mathieu equation~\cite{2018_Goldstein}. The Floquet analysis for this equation reveals that the natural frequencies of the pendulum can resonate with the driving at integer (harmonic) or half-integer (subharmonic) multiples of the driving frequency~\cite{1966_Mangus}. The essential symmetry-breaking phenomenon underlying discrete time crystals---the subharmonic response---has long been known to occur in many dissipative systems, including Faraday waves~\cite{1993_Cross_Hohenberg} and coupled-oscillator systems~\cite{2013_Braun}. However, it has been recently emphasized that both classical and quantum time crystals should also retain their essential structure when coupled to a heat bath in order to be considered a rigid phase of matter \cite{2018_Yao_Zaletel,2019_Gambetta,2019_Heugel,2020_Lazarides}. This implies that long-range order persists regardless of system imperfections and small external influences out to exponentially long time scales, much in the same way that ordinary phases of matter maintain their essential properties despite imperfections and external influences.

While most previous time crystals exhibit a partially broken time-translational symmetry that is commensurate with the driving in the form of the subharmonic response, other responses have also been noted. These responses include, for example, choreographic time crystals \cite{2020_Libal}, fractional frequency responses \cite{2019_Matus}, and coherent responses to quasiperiodic driving \cite{2018_Dumitrescu}. Incommensurate quantum time crystals, which exhibit neither harmonic nor subharmonic responses under periodic driving but maintain coherence in the form of quasiperiodicity, have been observed in periodically driven magnon condensates \cite{2018_Autti} and ultracold atoms \cite{2018_Giergiel,2019_Giergiel,2019_Cosme,2019_Pizzi}. Incommensurate responses have also been described in few-body classical systems \cite{1963_Hsu,1968_Yamamoto,1985_Hansen} and symbolic dynamics \cite{2018_Flicker}. However, classical discrete time crystals with incommensurate responses have not been described, and it remains unclear by which mechanisms an incommensurate response could emerge in the classical regime. 

In this paper, we report on a general mechanism giving rise to classical discrete time crystals with incommensurate frequency responses. This mechanism is based on a symmetry-breaking phenomenon in which an incommensurate response emerges directly from the symmetry-unbroken phase, which we call the \textit{anharmonic response}. Unlike traditional Hopf bifurcations that arise from the excitation of a single mode of instability, anharmonic responses arise from the excitation of a pair of modes that coresonate with the driving, resulting in a bifurcation between a fixed point and an invariant torus with no intermediate limit cycles.  We study the properties of the resulting anharmonic time crystal phase and characterize conditions under which it can emerge.

We proceed in Sec.~\ref{pendsec} by introducing a model system consisting of coupled pendula with alternating lengths. In Sec.~\ref{linsec}, we develop the Floquet theory that characterizes this model's response to parametric driving for zero temperature. In Sec.~\ref{ansec} , we then demonstrate anharmonic responses in this model and describe how they emerge in the Floquet theory through the coresonance of wave modes in different branches of the dispersion relation. In Sec.~{\ref{patternsec}, we consider the effects of positive temperature and the formation of patterns in large arrays of pendula. In Sec.~}\ref{gensec}, we generalize the mechanism to a broad class of weakly damped media composed of coupled fields. We conclude in Sec.~\ref{discussion} with a discussion of theoretical implications and potential applications of anharmonic responses.

\section{Anharmonic pendulum model}
\label{pendsec}
\begin{figure}[t]
\includegraphics[width=0.95\columnwidth]{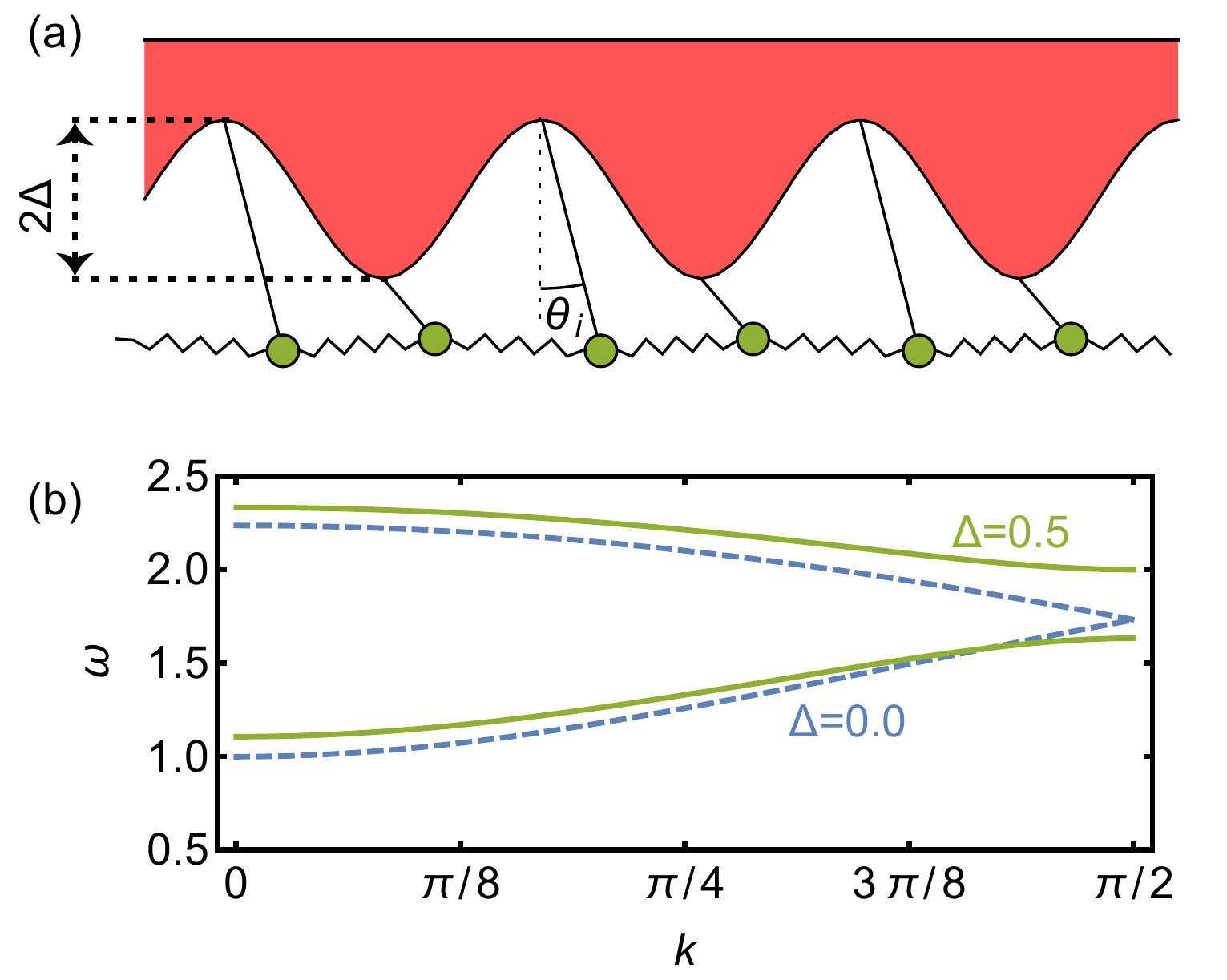}
\caption{Model system exhibiting an anharmonic response to periodic driving. (a) Array of coupled pendula of alternating lengths, which can oscillate about their pivots when driven by vertical vibrations through an anharmonic response incommensurate with the driving. (b) Dispersion relation between {the angular} frequency $\omega$ and wave number $k$, which governs wave propagation in the undriven pendulum array for $\Delta=0$ (blue dashed lines) and $\Delta=0.5$ (green solid lines). For $\Delta>0$, a band gap opens between the low-frequency (acoustic) modes and high-frequency (optical) modes and is harnessed to produce the anharmonic response. The anharmonic response is illustrated dynamically in the Supplemental Video~\cite{SM}, along with an animated summary of the main results.
\label{fig1}}
\end{figure}

In our anharmonic time crystal model, individual pendula experience gravitational forces $Mg$ and are coupled to nearest neighbors via linear springs with identical spring constants $\kappa$ [Fig.~\ref{fig1}(a)]. The springs have zero unstretched length and act with an attractive force proportional to the two-dimensional displacement between the particles. The pendulum rods are assumed to be massless and of fixed length. Crucially, the lengths of the pendula $L_i$ alternate in our model as $L_i = {l}+(-1)^i \Delta$, giving rise to a band gap in the dispersion relation [Fig.~\ref{fig1}(b)]{, which separates low-frequency (acoustic) modes from high-frequency (optical) modes}. The pendula are driven by the vertical sinusoidal motion of the supporting ceiling, which oscillates with a frequency $\omega_d$ and an amplitude $A_d$. For strong driving, the pendula begin to swing about their pivots in response.

We assume that in addition to the nearest-neighbor interactions, the pendula are coupled to a heat bath that is maintained at a temperature $T$. This coupling induces a dissipative force $-\eta L_i \dot{\theta}_i$ as well as a fluctuating, random force $\xi_i(t)$ on each pendulum, which is assumed to be Gaussian, white, and uncorrelated across $i$. Thus, $\langle \xi_i(t) \xi_j(t') \rangle = \sigma^2 \delta(t-t')\delta_{ij}$, where $\delta(t-t')$ is the Dirac delta and $\delta_{ij}$ is the Kronecker delta. According to the fluctuation-dissipation theorem, the noise intensity $\sigma$ and the dissipation strength $\eta$ are related according to $T=\sigma^2/2\eta$. In the oscillating reference frame of the supporting ceiling, the equations of motion are then
\begin{align}
\label{springs}
M L_i \ddot{\theta}_i &= -\eta L_i \dot{\theta}_i-M \left[g {\ +\ }  A_d \omega_d^2 \cos(\omega_d t)\right]\sin(\theta_i) \nonumber \\
& \quad +\kappa L_{i+1} \sin(\theta_{i+1}-\theta_i) +\kappa L_{i-1} \sin(\theta_{i-1}-\theta_i) \nonumber \\
&\quad + \kappa\left(L_{i+1}+L_{i-1}-2L_i\right) \sin(\theta_i) + \xi_i,
\end{align}
where the overdots denote derivatives with respect to time $t$ and $\eta$ is a damping coefficient. We consider the limit of infinitely many pendula but carry out numerical analysis for a finite number of pendula $N$ with periodic boundary conditions, in both cases for $\Delta=0.5$ unless otherwise noted. We nondimensionalize all variables with length and time scales such that $M=1$ and ${l}=1$, and{, for concreteness, we fix the non-dimensional gravity and damping strength to} $g=1$ and $\eta=0.1$ throughout. In the limit of small $\theta_i$, our model corresponds to the linearly coupled Frenkel-Kontorova model~\cite{2013_Braun}. {The initial growth of instabilities and small-amplitude wave dispersion can be predicted from the Frenkel-Kontorova model or the dispersion relation in this limit, but accurate prediction of nonlinear saturation of instabilities requires the inclusion of the fully nonlinear sinusoidal coupling terms in Eq.~\eqref{springs}}. 

\section{Floquet theory for parametric driving at $T=0$ }
\label{linsec}
Since the system in Eq.~\eqref{springs} is translationally invariant with respect to $i\to i+2$ when $T=0$, we can decouple waves in the linear regime using the Fourier ansatz
\begin{equation}
\theta_i =
 \begin{cases}
e^{\mathrm{i} ki}\phi_0(k) & \text{for even $i$}, \\
e^{\mathrm{i} ki}\phi_1(k) & \text{for odd $i$},
\end{cases}
 \label{fourier}
\end{equation}
where $\mathrm{i}$ is the imaginary unit. Separate Fourier amplitudes $\phi_0(k)$ and $\phi_1(k)$ are required for the short and long pendula in each unit cell in the lattice. General states can be expressed as a superposition of such wave modes over $k=i\pi/N$ for $i=0,1,\cdots\!,N/2$. Inserting Eq.~\eqref{fourier} into Eq.~\eqref{springs} results in a pair of coupled Mathieu equations for each wave number $k$,
\begin{align}
\label{springs2}
M ({l}+\Delta) \ddot{\phi}_0 &= -\eta ({l}+\Delta)\dot{\phi}_0-M \left[g+A_d \omega_d^2 \cos(\omega_d t)\right]\phi_0 \nonumber \\
& \quad + {2\kappa ({l}-\Delta) \cos(k)}\phi_1 -2 \kappa({l}+\Delta)\phi_0, \\
M ({l}-\Delta) \ddot{\phi}_1 &= -\eta({l}-\Delta) \dot{\phi}_1-M \left[g+A_d \omega_d^2 \cos(\omega_d t)\right]\phi_1 \nonumber \\
& \quad + {2\kappa ({l}+\Delta) \cos(k)}\phi_0 -2 \kappa({l}-\Delta)\phi_1, 
\label{springs3}
\end{align}
where we have suppressed the dependence of $\phi_0$ and $\phi_1$ on the wave number $k$ for brevity. Similar coupled Mathieu equations have been previously studied in contexts outside of many-body dynamical phases~\cite{1963_Hsu,1968_Yamamoto,1985_Hansen}.

We take advantage of the periodicity of the driving force with the Floquet ansatz
\begin{align}
\label{floquet}
\phi_i &= e^{s\omega_d t}\sum_m {\Phi} _{i m} e^{ \mathrm{i} m \omega_d t},
\end{align}
where $s=-\beta + \mathrm{i} \epsilon$ is the Floquet exponent, with scaled decay rate $\beta$ and response frequency ratio $\epsilon$, which are to be determined as functions of $k$. Just as the Fourier ansatz ensures that the modes form a representation of the space-translational symmetry group, the Floquet ansatz in Eq.~\eqref{floquet} ensures that the modes form a representation of the time-translational symmetry group. Below, we consider the quantity {$\omega_r \equiv \epsilon \omega_d$} for $\epsilon$ taken in the ``first Brillouin zone,'' which we call the dominant response frequency. The dominant response frequency specifies only a single frequency component of the total response in Eq.~\eqref{floquet}, which also contains frequency components $(\epsilon+m)\omega_d$ for all integers $m$. Importantly, since the sum in Eq.~\eqref{floquet} can always be redefined by a shift $m\to m+1$, the ratio $\epsilon$ is only defined up to $\mathrm{mod}~1$ congruences. This means that, for example, $\epsilon=\cdots,-1/2,1/2,3/2,\cdots$ are all equivalent ways to describe subharmonic responses occurring at half the driving frequency while $\epsilon=\cdots,0,1,2,\cdots$ all correspond to harmonic responses occurring at the driving frequency. We focus primarily on the first Brillouin zone defined by $0 < \epsilon < 1$, so that $\epsilon \omega_d$ is the positive frequency component of smallest magnitude in Eq.~\eqref{floquet}, but we also use $-1/2 < \epsilon < 1/2$ when convenient.

It should be noted that there is a complex conjugate solution $-\beta-\mathrm{i}\epsilon$ associated with each Floquet exponent $-\beta+\mathrm{i}\epsilon$ for positive $\epsilon$. This solution contains the frequency components $(-\epsilon+m)\omega_d$ for all integers $m$. Thus, taken together, the frequency ratios in the first Brillouin zone are $\epsilon$ and $1-\epsilon$ for each acoustic and each optical mode.

Substitution of the Floquet ansatz in Eq.~\eqref{floquet} into Eqs.~\eqref{springs2} and \eqref{springs3} results in the infinite-dimensional quadratic $s$-eigenvalue problem
\begin{equation}
\label{eigen1}
s^2 \sum_{i m} A^{i m}_{j n} {\Phi} _{i m} + s \sum_{i m} B^{i m}_{j n} {\Phi} _{i m} + \sum_{i m} C^{i m}_{j n} {\Phi} _{i m}=0,
\end{equation}
{where $0\leq i,j \leq1$ and $-\infty<n,m<\infty$ are integers. Here,}
\begin{align}
A^{i m}_{j n} &= {L_i}M\omega_d^2\delta^i_j\delta^m_n, \label{amat} \\
B^{i m}_{j n} &= {L_i}\left( {2\mathrm{i}}M\omega_d^2n+\omega_d\eta \right)\delta^i_j\delta^m_n, \label{bmat} \\
C^{i m}_{j n} &= {L_i}\left( -M\omega_d^2n^2+\mathrm{i}\omega_d\eta n+2\kappa \right)\delta^i_j\delta^m_n \label{cmat} \nonumber \\
&\quad 
-{2\kappa L_i \left( \delta^i_{j+1} + \delta^i_{j-1}\right)} \cos (k) \delta^m_n \nonumber \\
&\quad + M g\delta^i_j\delta^m_n + \frac{1}{2}MA_d\omega_d^2\left(\delta^m_{n+1} + \delta^m_{n-1}\right)\delta^i_j,
\end{align}
where $\delta^i_j$ is the Kronecker delta {and we have reintroduced $L_i=l+(-1)^i\Delta$ to simplify the expressions}. 
Linearization of this nonlinear eigenvalue problem \cite{2001_Tisseur_Meerbergen} can be achieved by extending the system to
\begin{equation}
\sum_{im}
\mbox{$\begin{pmatrix*}
0 & \delta^{im}_{jn} \\
C^{im}_{jn} & B^{im}_{jn}
\end{pmatrix*}$}
 {\mbox{$\begin{pmatrix*}
\Phi_{im} \\
\zeta_{im}
\end{pmatrix*}$}} =
s \sum_{im}\mbox{$\begin{pmatrix*}[r]
\delta^{im}_{jn} & 0 \\
0 & -A^{im}_{jn}
\end{pmatrix*}$}
{\mbox{$\begin{pmatrix}
\Phi_{im} \\
\zeta_{im}
\end{pmatrix}$}} ,
\end{equation}
where $\zeta_{im}$ are auxiliary variables and $\delta^{im}_{jn}=\delta^i_j\delta^m_n$.
In our subsequent analysis, the eigenvalues for this system are found numerically by truncating beyond Floquet modes $-5\leq n,m \leq 5$. When the decay rate $\beta$ is negative for some wave number $k$, the pendula resonate with the driving force and begin to swing, while they remain motionless in the oscillating reference frame when $\beta>0$ for all $k$.
\begin{figure*}
 \includegraphics[width=2\columnwidth]{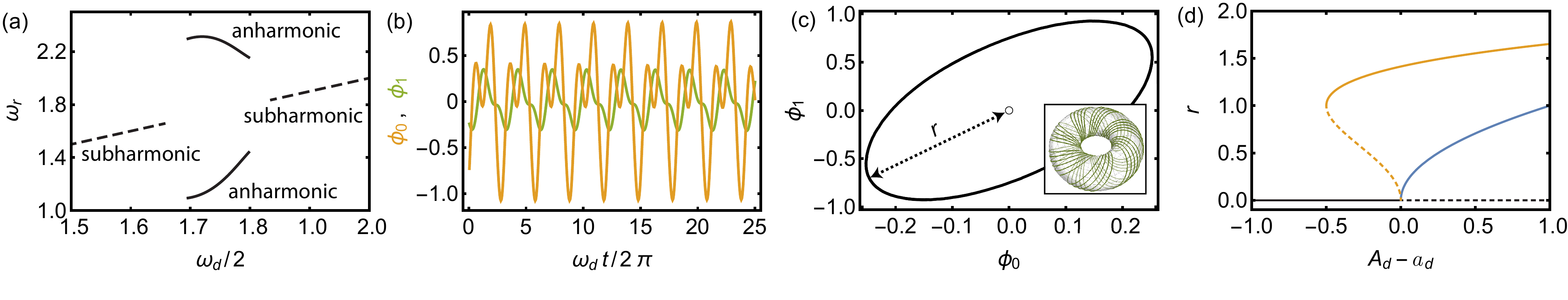}
 \caption{Anharmonic responses in the time crystal model. (a) Dominant response frequencies {$\omega_r$} vs.\ driving frequency $\omega_d/2$ for unstable $k=0$ modes in the pendulum array. The anharmonic response manifests as a pair of solid curves, corresponding to $\epsilon$ and $1-\epsilon$. (b) Phases $\phi_0$ and $\phi_1$ vs.\ normalized time for the saturated anharmonic instability corresponding to $\omega_d/2=1.7$. The system exhibits approximately $8$ periods as the driving force completes $25$ periods. (c) Orbit of the stroboscopic map in Eq.~\eqref{strobe}, with the unstable fixed point ({open circle}) corresponding to the non-swinging state and the stable invariant curve (solid curve) that emerges from the Neimark-Sacker bifurcation. The invariant curve in the stroboscopic map corresponds to an invariant torus (inset) in the continuous time dynamics in Eq.~\eqref{springs}, giving rise to the anharmonic response. (d) Schematic of the response amplitude $r$ vs.\ the driving amplitude $A_d-a_d$ for the stable (solid lines) and unstable (dashed lines) solution branches. The nonswinging stationary state (black lines) loses stability through interactions with invariant torus solution branches, which may have a subcritical (orange lines) or supercritical (blue lines) transition. \label{fig2}}
\end{figure*}

From a different perspective, the instability boundaries corresponding to the onset of instability can be identified by constraining the decay rate in the Floquet exponents to $\beta=0$. The linear system in Eq.~\eqref{eigen1} can be reinterpreted in terms of the $a_d$-eigenvalue problem
\begin{equation}
\label{eigen2}
\sum_{i m} D^{i m}_{j n} {\Phi} _{i m} = a_d\sum_{i m} E^{i m}_{j n} {\Phi} _{i m},
\end{equation}
where
\begin{align}
D^{i m}_{j n} &= {L_i}\left[ M\omega_d^2(s+{\mathrm{i}}n)^2+\omega_d\eta(s+{\mathrm{i}}n)+2\kappa \right]\delta^i_j\delta^m_n \nonumber \\
 &\quad -{2\kappa L_i\left(\delta^i_{j+1} + \delta^i_{j-1}\right)}\cos (k) \delta^m_n + M g\delta^i_j\delta^m_n , \\
E^{i m}_{j n} &= {-}\frac{1}{2}M\omega_d^2\left(\delta^m_{n+1} + \delta^m_{n-1}\right)\delta^i_j.
\end{align}
The real eigenvalue solutions $a_d$ to Eq.~\eqref{eigen2} for values of the Floquet exponent $s=0+\mathrm{i} \epsilon$ then correspond to the values of $A_d$ on the stability boundary for an instability with dominant frequency ratio $\epsilon$. The subharmonic and harmonic instability boundaries can be mapped out by setting $\epsilon=0$ and $\epsilon=1/2$ in Eq.~\eqref{eigen2}, but the value of $\epsilon$ generally varies with $\omega_d$ for the anharmonic instability boundaries.

\section{Emergence of anharmonic responses}
\label{ansec}
Figure \ref{fig2}(a) shows how the dominant response frequencies of unstable modes determined from Eq.~\eqref{eigen1} vary with the drive frequency $\omega_d$ for fixed $A_d=0.05$. The anharmonic response appears in the range $1.7 \lesssim \omega_d \lesssim 1.8$, where two branches of unstable modes appear. Since the anharmonic response frequencies vary continuously (and even non-monotonically), the ratio $\epsilon$ is typically irrational and the modes corresponding to $\epsilon$ and $1-\epsilon$ are thus incommensurate with each other, implying that the aggregated response from these modes is quasiperiodic. In direct numerical simulations (see Appendix \ref{numericsec}), the growth of the instability does not continue indefinitely but instead saturates outside the linear regime due to nonlinear effects. The saturated response is a coherent oscillation of finite amplitude, as shown in Fig.~\ref{fig2}(b) for the phases determined by Eqs.~\eqref{springs2} and \eqref{springs3}.

The bifurcation leading to the anharmonic response can be classified by considering the stroboscopic map
\begin{equation}
{\label{strobe}
\left[\phi_0(t),\phi_1(t)\right]\to \left[\phi_0(t+\omega_d/2\pi), \phi_1(t+\omega_d/2\pi)\right],}
\end{equation}
which is obtained by strobing the system at the driving frequency. The non-swinging state in the stroboscopic map corresponds to a fixed point at the origin. For driving amplitudes above the critical driving amplitude $a_d$ in Eq.~\eqref{eigen2}, this fixed point is unstable, and the system is attracted to a different invariant set. For the subharmonic response, this attractor corresponds to a period-$2$ orbit that emerges from the fixed point via a period-doubling bifurcation. For the anharmonic response, on the other hand, the attractor is an invariant curve that emerges from a Neimark-Sacker bifurcation \cite{1979_Iooss}, as shown in Fig.~\ref{fig2}(c). The Neimark-Sacker bifurcation is {perhaps best} known for describing the Poincar\'{e} map of a torus bifurcation, in which a limit cycle (corresponding to the fixed point in the map) interacts with a quasiperiodic invariant torus (corresponding to the invariant curve in the map). Incommensurate responses {were previously found} to emerge via a Hopf bifurcation, which generates a limit cycle, followed by a secondary torus bifurcation, which generates an invariant torus from the limit cycle.  Importantly, since the origin in the stroboscopic map in Fig.~\ref{fig2}(c) corresponds to a fixed point rather than a limit cycle,  the anharmonic response emerges from an entirely different bifurcation involving a fixed point and an invariant torus with no intermediate limit cycle. Such a bifurcation is structurally unstable for autonomous systems, but as we have shown, this bifurcation is possible for periodically driven systems.

The Neimark-Sacker bifurcation can take either a supercritical or subcritical form, as illustrated in Fig.~\ref{fig2}(d). The criticality of the bifurcation depends on nonlinear terms in Eq.~\eqref{springs} beyond the linear approximation in Eqs.~\eqref{springs2} and \eqref{springs3}. Subcritical bifurcations correspond to discontinuous transitions, which exhibit hysteresis, while supercritical transitions correspond to continuous and reversible transitions. We can determine the bifurcation criticality numerically. To do so, we allow the instability to grow until it saturates to the nonlinear steady-state with $A_d>a_d$. Then, we quasistatically decrease the driving amplitude to $A_d<a_d$ and observe whether the system returns to the non-swinging state or exhibits hysteresis. {In principle, the frequency ratio and wave number of the saturated response can differ substantially from that predicted by the linear dispersion relation for subharmonic bifurcations, as they can undergo large variations in their departure along the unstable branch. Numerically, however, we find that  these quantities do not deviate from the linear dispersion relation predictions in the pendulum array because the subcriticality is never largely pronounced in this case.} 

 \begin{figure}[t]
\includegraphics[width=\columnwidth]{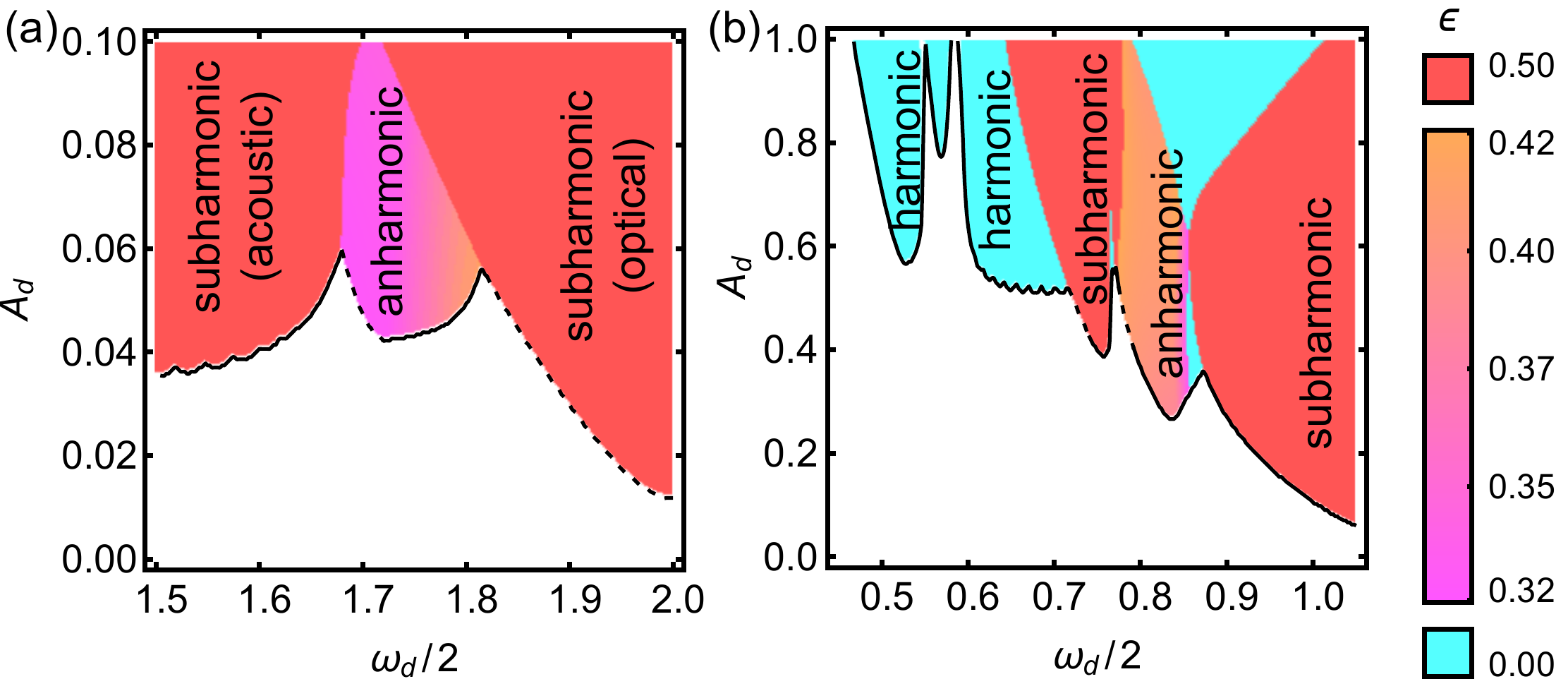}
\caption{Parametric instabilities in the anharmonic time crystal model. (a) Frequency ratio $\epsilon$ for the most prominent subharmonic and anharmonic instability tongues. The anharmonic instability emerges between the acoustic and optical subharmonic instabilities with response frequency ratios $\epsilon$ and $1-\epsilon$ for $0.32<\epsilon<0.42$. The instabilities leading to the anharmonic and subharmonic responses are subcritical for ranges of driving frequencies delimited by $\omega_d/2<1.72$ and $\omega_d/2<2.0$, respectively, as indicated by the dashed line, and supercritical elsewhere. (b) Frequency ratio $\epsilon$ for less prominent tangles of harmonic, anharmonic, and subharmonic responses at lower frequencies and higher driving amplitudes. \label{fig3}}
\end{figure}
\begin{figure*}
\includegraphics[width=1.9\columnwidth]{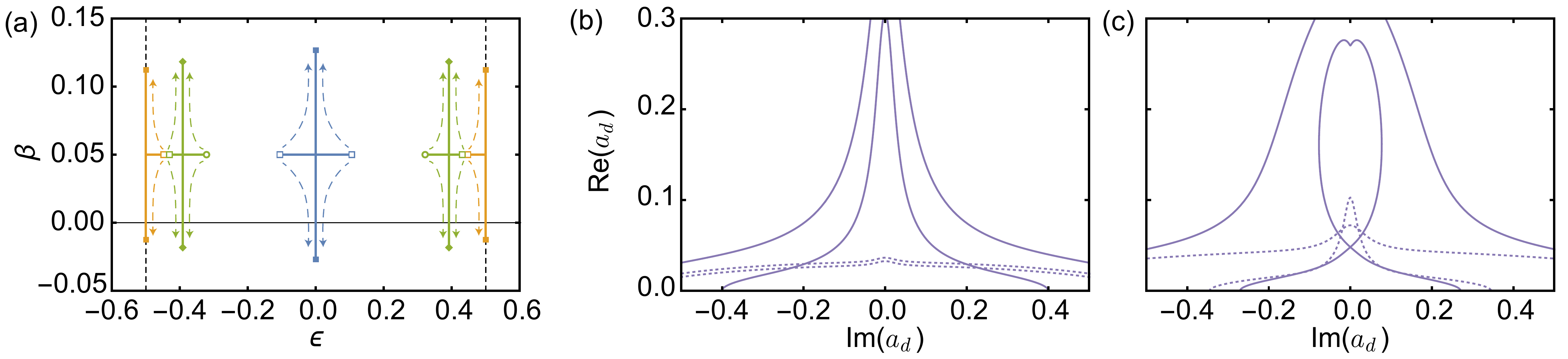}
\caption{Floquet analysis of the anharmonic time crystal model. (a) Floquet exponents $s=-\beta+\mathrm{i}\epsilon$ in the complex plane for harmonic (blue), subharmonic (orange), and anharmonic (green) instabilities. The open (closed) symbols show the exponents for driving forces below (above) the instability boundary in Fig.~\ref{fig3}. The solid lines show the trajectories of the exponents as $A_d$ increases, with dashed lines and arrows to guide the eye on the direction of the motion. Squares correspond to acoustic modes, circles to optical modes, and diamonds to coalesced acoustic and optical modes. (b), (c) Eigenvalues $a_d$ corresponding to the critical driving amplitude for modes with $k=0$ (solid lines) and $k=\pi/2$ (dotted lines) for $\Delta=0$ (b) and $\Delta=0.5$ (c) as $\epsilon$ varies from $-0.5$ to $0.5$ for ${\omega_d=3.4}$. \label{fig4}}
\end{figure*}

Figure \ref{fig3} shows the frequency ratio $\epsilon$ for the unstable modes with the largest growth rate $-\beta$ in the driving amplitude vs.\ driving frequency space, as determined by the Floquet analysis defined by Eq.~\eqref{eigen1}. The wave number of the dominant instability varies within each instability tongue. In the anharmonic tongues, on the other hand, not only the wave number but also the frequency ratio varies continuously. Because of the band gap that opens between the acoustic and optical modes of the dispersion relation [Fig.~\ref{fig1}(b)], there is a separation between the corresponding subharmonic instabilities in the driven array [Fig.~\ref{fig3}(a)]. Remarkably, the anharmonic instability tongue emerges precisely in this band gap and, depending on the driving frequency, the instability can be either supercritical or subcritical. For driving frequencies higher than those shown in Fig.~\ref{fig3}(a), the response is dominated by short-wavelength subharmonic instabilities, while for lower frequencies, a tangle of additional harmonic, anharmonic, and subharmonic instabilities appears [Fig.~\ref{fig3}(b)]. For driving amplitudes far above the stability boundaries in Fig.~\ref{fig3}, multiple modes are simultaneously excited, which generally results in a less rigid, chaotic response that lacks any time crystallinity. Incidentally, in other systems, interesting localized and topological states have also been shown to occur around band gaps~\cite{2004_Cambell, 2016_PonedeL_Knobloch,2018_MitchelL_Irvine}.

We now examine the emergence of anharmonic instabilities in terms of the interactions between Floquet exponents in the anharmonic time crystal model. Figure \ref{fig4}(a) shows how Floquet exponents come together to produce instabilities as the driving amplitude increases for harmonic, subharmonic, and anharmonic instabilities. For harmonic instabilities (blue symbols and lines), the positive and negative frequency components of a single acoustic mode of the dispersion relation in Fig.~\ref{fig1}(b) coalesce along the horizontal direction at $\epsilon=0$ and separate along the vertical direction to produce instability. For the subharmonic instabilities (orange symbols and lines), a similar process occurs at the edge of the Brillouin zone ($\epsilon=\pm 0.5$). In both cases, depending on $\omega_d$, the instability can instead be produced by an optical mode. Thus, the harmonic and subharmonic instabilities occur generically because Floquet exponents appear as complex conjugates, singling out the values $\epsilon=0$ and $\epsilon=\pm 0.5$ at which negative and positive frequency modes can coalesce. For the anharmonic instabilities (green symbols and lines), on the other hand, the acoustic and optical modes interact with each other at a value of $\epsilon$ between $0$ and $\pm 0.5$ before moving apart along the vertical direction $\beta$. These anharmonic instabilities therefore correspond to the coresonances that can occur more generally at sums and differences of the natural (undriven) frequencies in the coupled Mathieu equations. The peaks in the {Fourier transform of the signal} in Fig.~\ref{fig2}(b) correspond to the frequencies predicted by the Floquet exponents in Fig.~\ref{fig4}(a) ({i.e., the} $\epsilon$ at which the green modes coalesce), confirming that this coresonance gives rise to the saturated anharmonic response.

For specified harmonic ($\epsilon=0$) or subharmonic ($\epsilon=\pm 0.5$) frequency ratio, the instability boundary can be derived from the smallest real solution of the $a_d$ eigenvalue of the problem in Eq.~\eqref{eigen2}. For anharmonic instabilities, on the other hand, we must allow $\epsilon$ to vary continuously between $-0.5$ and $0.5$ and detect the minimal values of $\mathrm{Re}(a_d)$ for which $\mathrm{Im}(a_d)=0$ to find the instability boundary at $A_d=\mathrm{Re}(a_d)$. Figure \ref{fig4}(b) shows the resulting eigenvalues for the modes in Eq.~\eqref{fourier} with $k=0$ and $k=\pi/2$ when $\Delta=0$. In this case, the $k=\pi/2$ mode goes unstable first (i.e., at lower $A_d$). The symmetry in the system implies that the $\epsilon=0$ and $\epsilon=\pm 0.5$ values always correspond to purely real or imaginary $a_d$, but additional real eigenvalues can emerge when $\Delta>0$ at self-intersecting loops in the eigenvalue traces. These solutions can overtake the $k=\pi/2$ mode (by occurring at a lower $A_d$) as $\Delta$ increases, as shown in Fig.~\ref{fig4}(c) for the $k=0$ mode when $\Delta=0.5$. Direct numerical simulations {also} confirm that these loop intersections correspond  to the boundary of the anharmonic instabilities in Fig.~\ref{fig3}.

\section{{Pattern formation for $T>0$}}
 \label{patternsec}
We now consider the patterns that develop in large arrays of pendula at finite temperatures. While the noise terms in Eq.~\eqref{springs} break the instantaneous translational symmetry $i \to i+2$ for $T>0$, a Floquet-Fourier analysis is possible for the distribution of phases in ensembles of systems at finite temperatures through the corresponding Fokker-Planck equations. Such an analysis is beyond the scope of this work, however, and instead we investigate the impact of temperature through numerical simulations.

\begin{figure*}
\includegraphics[width=2\columnwidth]{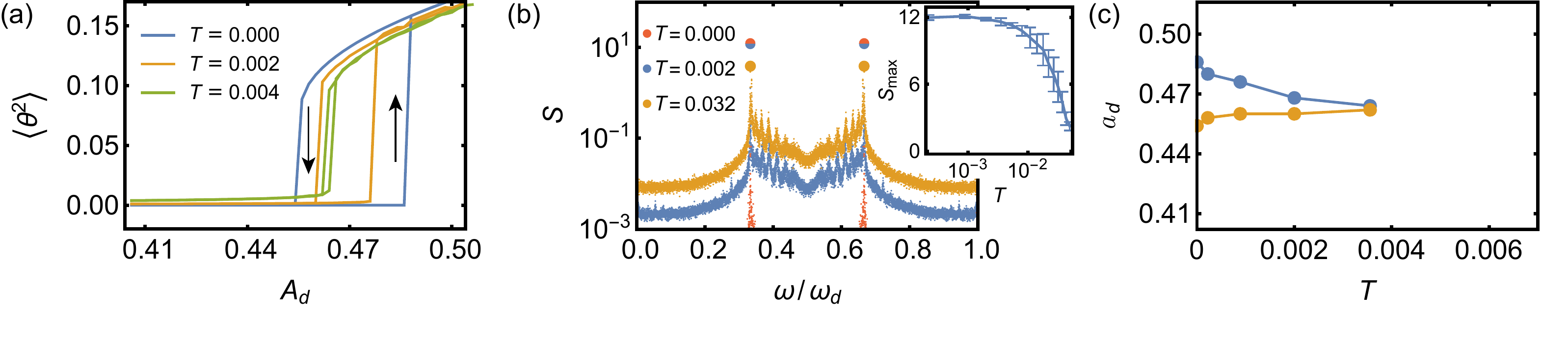}
\caption{Discontinuous transition for the anharmonic response at $\omega=3.4$. (a) Time-averaged pendulum angle $\langle \theta^2 \rangle$ vs.\ the driving amplitude $A_d$ under quasistatic variation, exhibiting hysteresis which only vanishes for sufficiently large $T$. (b) Spectral power $P$ vs.\ frequency $\omega$ for the anharmonic response in the absence and presence of noise. The prominent peaks (large dots) are unaffected in position ($\epsilon\approx 0.32$ and $1-\epsilon \approx 0.68$) and magnitude ($S=S_{\mathrm{max}}$) by small noise ($T\lesssim 0.004$, overlapping blue and red dots), but the magnitude begins to decrease for larger temperatures (orange dots). The magnitude of the peaks as a function of the temperature is plotted in the inset. (c) Upper (blue) and lower (orange) critical driving driving amplitudes $a_d$ vs.\ the temperature $T$ for the {subcritical} anharmonic transition. The hysteretic transition persists for small $T$ but is smoothed out as the upper and lower critical amplitudes coalesce at sufficiently large $T$, coinciding with the decay in the peaks of the spectral power. \label{fig5}}
\end{figure*}

Crucially, our simulations indicate that the anharmonic response at zero temperature is a result of a discontinuous, subcritical transition for $1.68\lesssim \omega_d/2 \lesssim 1.72$, as shown in Fig.~\ref{fig5}(a). While supercritical transitions are smeared out by arbitrarily small temperatures, subcritical transitions are structurally stable against the impact of small temperatures. This stability against the impact of finite temperatures is reflected in the spectral power of the saturated response, as shown in Fig.~\ref{fig5}(b). The power spectrum here is the average magnitude of the discrete Fourier transform of the stroboscopic map in Eq.~\eqref{strobe}. With or without noise, the response is composed of two modes with dominant frequency components, $\epsilon$ and $1-\epsilon$, which are not half-integer multiples of the driving frequency. As in Eq.~\eqref{floquet}, there are also less prominent frequency components at $(\epsilon+m)\omega_d$ and $(1-\epsilon+m)\omega_d$ for all integer $m$. Crucially, the peaks position and magnitude are not significantly affected by small noise ($T\lesssim0.004$) but instead remain rigid. Only for larger temperatures does the magnitude of the peaks begin to decay and is the order in the phase lost, as shown in the inset. Indeed, the peaks begin to decay for temperatures sufficiently large that the hysteresis loop disappears and the transition between the swinging and non-swinging states becomes smooth, as shown in Fig.~\ref{fig5}(c). Thus, we confirm that the subcriticality of the transition grants the anharmonic phase a degree of rigidity against the influence of finite temperature. The same rigidity is seen when spatial disorder is introduced in the form of small random perturbations to the lengths of the pendula. The anharmonic response is therefore fundamentally distinct from the splitting of a subharmonic response due to experimental imperfections such as nonperiodic driving \cite{2017_Zhang,2017_Yao}.
\begin{figure}[b!]
\includegraphics[width=0.95\columnwidth]{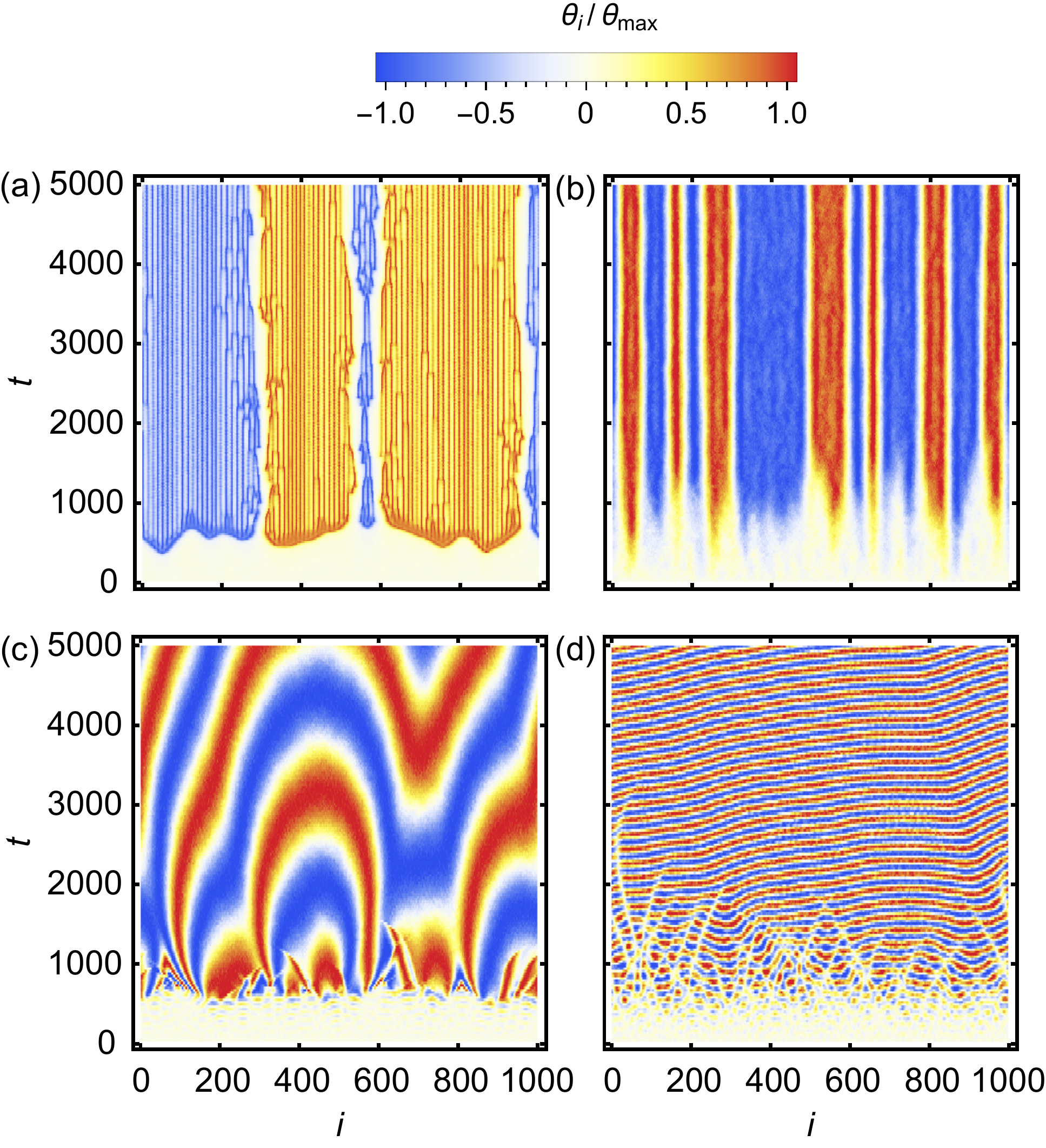}
\caption{Spatiotemporal evolution of patterns in the pendulum array, where $i$ indexes the individual pendula. (a), (b) Phase defects in subharmonic responses for $\omega_d=3.90$ and $a_d=0.019$ corresponding to a subcritical transition (a) and for $\omega_d=4.00$ and $a_d=0.013$ corresponding to a supercritical transition (b). (c), (d) Waves of phase variation in anharmonic responses for $\omega_d=3.40$ and $a_d=0.050$ corresponding to a subcritical transition (c) and for $\omega_d=3.44$ and $a_d=0.045$  corresponding to a supercritical transition (d). The color indicates the amplitude of the swinging pendulum angles (strobed at frequencies commensurate with the dominant response frequencies) relative to the maximum $\theta_{\mathrm{max}}$ for each case. \label{fig6}}
\end{figure}

In large pendulum arrays ($N=1000$), the spatiotemporal order slowly varies, creating patterns of variation in the arrays. Figure \ref{fig6} shows how these patterns differ significantly depending on the type of bifurcations leading to the instabilities. As in previous studies \cite{2018_Yao_Zaletel}, the subharmonic instability is dominated by alternating domains [Figs.~\ref{fig6}(a) and \ref{fig6}(b)], which are swinging with two possible phases relative to the driving. Defects separate these domains, which are stationary in the supercritical case and motile in the subcritical case. The patterns are entirely different for the anharmonic instabilities [Figs.~\ref{fig6}(c) and \ref{fig6}(d)], since there is a continuum of phases available relative to the driving in this case. Slow variations in the phase give rise to Goldstone-like modes with long-wavelength patterns, which are again significantly more mobile in the subcritical case than in the supercritical case.

For traditional instabilities involving a finite number of unstable modes, the criticality of a bifurcation can be determined from the sign of nonlinear coefficients in the amplitude equations derived from weakly nonlinear analysis \cite{1993_Cross_Hohenberg}. However, since the instabilities here involve frequency components $(\pm\epsilon+m)\omega_d$ for all integer $m$, the weakly nonlinear analysis appears to involve infinitely many coupled amplitude equations, making this approach complicated. Although a complete weakly nonlinear description is beyond the scope of this work, previous studies on cellular automata suggest that the Kardar-Parisi-Zhang equation may give a qualitative description of these patterns \cite{Grinstein_1993,Chate_1995}. These patterns of defects and phase variations coarsen with time but persist indefinitely, ultimately destroying long range order and causing the spatiotemporal correlations to decay. Nevertheless, the structural stability of the discontinuous, subcritical transition giving rise to the anharmonic response represents a form of phase rigidity analogous to that of previous classical discrete time crystals \cite{2018_Yao_Zaletel}.

 \section{General conditions for coresonance}
\label{gensec}
We now generalize our results for $T=0$ to a wide class of systems beyond Eq.~\eqref{springs}. We consider a homogeneous and isotropic parametrically-driven extended medium modeled by a set of $\ell \geq2$ coupled fields $\theta_i$ for $i=0,1,\cdots,\ell-1$. Each $\theta_i$ represents a degree of freedom at each location in the medium, generalizing the displacements angles for the long and short pendula in the pendulum array. We assume that the $\theta_i$ evolve in space and time according to a set of coupled differential equations.

For weak parametric driving, the linear stability of the uniform state (described by $\theta_i=0$) characterizes the propagation of waves in the medium. We assume that, after linearizing and applying the Fourier transform to eliminate the spatial variables, the coupled equations take the form
\begin{align}
\label{coupled}
\frac{\partial^2 \phi_i}{\partial t^2} + \sum_j \left[ F^j_{i}(k) + A_d G^j_{i}(k)\cos(\omega_d t)\right]\phi_j = 0,
\end{align}
for $i,j=0,1,\ldots,\ell-1$, where $\phi_i$ is the Fourier transform of $\theta_i$, $k$ is the Fourier wave number, and $A_d$ and $\omega_d$ are the driving amplitude and frequency, respectively. [Here, we use partial time derivatives to emphasize the spatial dependence encoded by the wavenumber $k$.] Coupling is described by a coupling matrix $F$ with elements $F^j_{i}$ in the absence of driving, and the drive-induced coupling is described by another coupling matrix $G$ with elements $G^j_{i}$. Equation \eqref{coupled} represents a widely applicable second-order form that neglects damping terms for simplicity, but we also discuss modifications when we include weak first-order damping terms.

In the absence of driving ($A_d=0$), we denote the $j$th component {of the} $i$th eigenvector of $F$ by $\chi_{ij}$ and assume that the eigenvalues $\omega_{i}(k)^2$ are strictly positive, as required for a stable homogeneous state. Changing to the undriven eigenbasis given by $\psi_i$, where $\phi_j = \sum_{i} \psi_i \chi_{ij}$, Eq.~\eqref{coupled} becomes
\begin{align}
\label{diagonalized}
\frac{\partial^2 \psi_i}{\partial t^2} + \sum_j \left[ \omega_i(k)^2\delta^j_i+ A_d \widetilde{G}^j_{i}(k)\cos(\omega_d t)\right]\psi_j = 0,
\end{align}
with $\widetilde{G}^j_{i}(k) = \sum_{i'j'} \chi_{ij'} G^{j'}_{i'}(k) (\chi^{-1})^{i'j}$ for $(\chi^{-1})^{i'j}$ denoting the matrix elements of the inverse transformation to the eigenbasis.

The Floquet analysis in the undriven eigenbasis is carried out by eliminating the time dependence with the Floquet ansatz $\psi_i = e^{s \omega_d t}\sum_m \Psi_{im}e^{\mathrm{i}m \omega_d t}$, which transforms Eq.~\eqref{diagonalized} into
\begin{equation}
\label{genfloquet}
\sum_{jn} \Big[\omega_d^2(s+{\mathrm i}n)^2 \delta^{jn}_{im} + \omega_i(k)^2 \delta^{jn}_{im} + {A_d} \widehat{G}^{jn}_{im}(k)\Big] \Psi_{jn} = 0,
\end{equation}
where $n$ and $m$ range over all integers and $\widehat{G}^{jn}_{im}(k) = \widetilde{G}^j_{i}(k) \left(\delta^n_{m+1}+ \delta^n_{m-1}\right)/2$. Expressing Eq.~\eqref{genfloquet} in terms of $\mathbf{v}_{im} \equiv \begin{pmatrix}\Psi_{im} \\ s\Psi_{im}\end{pmatrix}$ linearizes the system, resulting in $\sum_{jn} H^{jn}_{im} \mathbf{v}_{jn} = s\mathbf{v}_{im}$, which is an eigenvalue problem for the Floquet exponents $s$ with matrix elements
\begin{equation}
\label{genmatrix}
H^{jn}_{im} =\begin{pmatrix} 0 & \delta^{jn}_{im} \\ \left[n^2-\omega_i(k)^2/\omega_d^2 \right] \delta^{jn}_{im} & -2\mathrm{i}n\delta^{jn}_{im} \end{pmatrix} + \frac{A_d}{\omega_d^2}\widehat{H}^{jn}_{im},
\end{equation}
where $\widehat{H}^{jn}_{im} = \begin{pmatrix} 0 & 0 \\ -\widehat{G}^{jn}_{im}(k) & 0 \end{pmatrix}$.

We proceed with a perturbative analysis of the eigenvalues of $H^{jn}_{im}$ for small $\lambda\equiv A_d/\omega_d^2$, where the perturbation is given by $\lambda \widehat{H}^{jn}_{im}$. By virtue of the diagonalization of the undriven system in the $\psi_k$ eigenbasis, the unperturbed eigenvalues are given by $s^{im\pm} \equiv -\mathrm{i}m \pm \mathrm{i}{\omega_i(k)}/{\omega_d}$, with corresponding eigenvectors $\mathbf{v}_{jn}^{im\pm} \equiv \begin{pmatrix}\delta^{jn}_{im} \\ s^{im\pm} \delta^{jn}_{im}\end{pmatrix}$. Since the eigenbasis determined by $\mathbf{v}_{jn}^{im\pm}$ is non-orthogonal, the perturbation theory should be carried out by projecting the eigenvalue problem onto a dual basis. The dual (row) vectors $\mathbf{w}_{im\pm}^{jn}$ corresponding to $\mathbf{v}_{jn}^{im\pm}$ must satisfy $\sum_{jn} \mathbf{w}_{im+}^{jn} \mathbf{v}^{i'm'+}_{jn} = \sum_{jn} \mathbf{w}_{im-}^{jn} \mathbf{v}^{i'm'-}_{jn}=\delta_{im}^{i'm'}$ and $\sum_{jn} \mathbf{w}_{im+}^{jn} \mathbf{v}^{i'm'-}_{jn} = \sum_{jn}\mathbf{w}_{im-}^{jn} \mathbf{v}^{i'm'+}_{jn}=0$, which implies that $\mathbf{w}_{im\pm}^{jn} = \mp\begin{pmatrix} s^{im\mp}\delta^{jn}_{im} & \delta^{jn}_{im} \end{pmatrix}/\left[2\mathrm{i}\omega_i(k)/\omega_d\right]$. In the non-degenerate case, on the one hand, the eigenvalues are unperturbed to first order in $\lambda$ since the perturbation term $\widehat{G}^{jn}_{im}(k)$ is zero when $n=m$. Given the strictly imaginary unperturbed eigenvalues, the second-order perturbation term is also imaginary and does not affect stability. Thus, to second order in $\lambda$, the driving does not induce any instabilities in the non-degenerate case. In the degenerate case, on the other hand, stability may be affected by driving at first or higher order in $\lambda$.
\begin{figure*}
 \includegraphics[width=1.9\columnwidth]{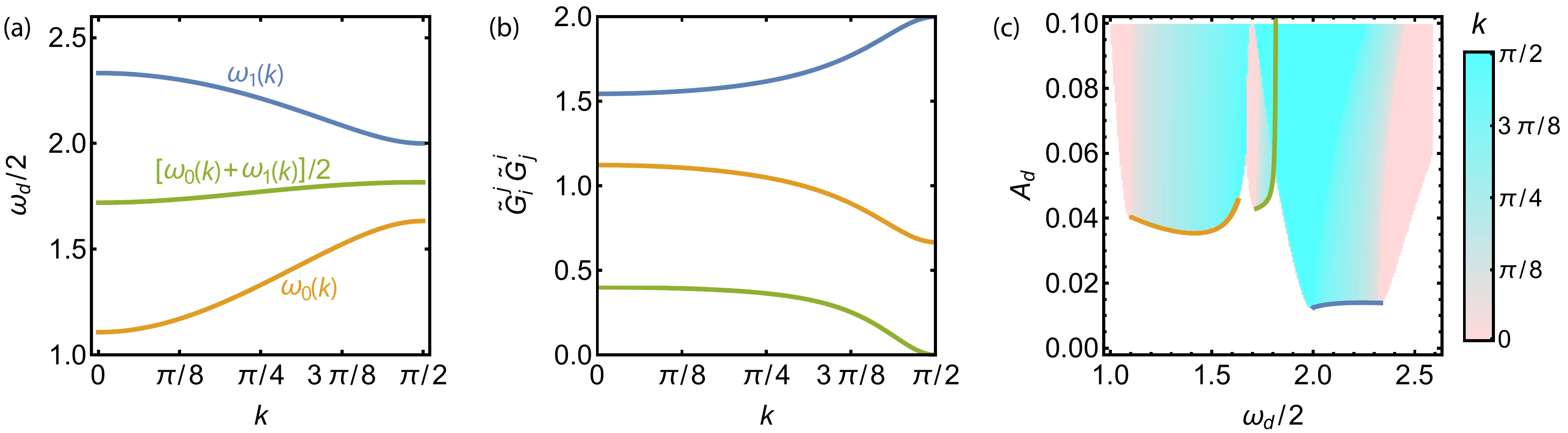}
 \caption{First-order degenerate theory for the anharmonic pendulum model. (a) Driving frequencies vs.\ the wave numbers for branches of coresonance that result in prominent instabilities. (b) Values of the drive-induced coupling matrix element product that determines the instability in Eq.~\eqref{instability} vs.\ the wave numbers for the color-corresponding resonance branches in (a). (c) Instability boundary determined by equating the first-order degenerate perturbation in Eq.~\eqref{instability} to the damping term $\eta/2\omega_d$ (solid lines), with colors corresponding to the resonance branches in (a). The colored shading shows instability regions corresponding to negative decay rates in non-perturbative solutions to Eq.~\eqref{eigen1} (which are in good agreement with direct numerical simulations), with the wave number of the most unstable mode indicated by the color bar. \label{fig7} }
\end{figure*}

The unperturbed eigenvalues $s^{im\pm}$ and $s^{jn\mp}$ become degenerate when
\begin{equation}
\label{resonance}
\omega_i(k) \pm \omega_j(k) = \pm (m-n)\omega_d,
\end{equation}
which specifies a coresonance condition between the driving and modes $i$ and $j$. The first-order perturbation of the eigenvalues in the degenerate case is determined by the eigenvalues of the matrix consisting of the degenerate space matrix elements of the perturbation
\begin{equation}
\label{degen}
{\cal G}~\equiv \sum_{i'm'j'n'}\begin{pmatrix} \mathbf{w}_{im\pm}^{i'm'}\widehat{H}_{i'm'}^{j'n'}\mathbf{v}^{im\pm}_{j'n'} & \mathbf{w}_{im\pm}^{i'm'}\widehat{H}^{j'n'}_{i'm'}\mathbf{v}^{jn\mp}_{j'n'} \\ \mathbf{w}_{jn\mp}^{i'm'}\widehat{H}^{j'n'}_{i'm'}\mathbf{v}^{im\pm}_{j'n'} & \mathbf{w}_{jn\mp}^{i'm'}\widehat{H}^{j'n'}_{i'm'}\mathbf{v}^{jn\mp}_{j'n'} \end{pmatrix}.
\end{equation}
 The matrix ${\cal G}$ in Eq.~\eqref{degen} is nonzero only for resonant modes in Eq.~\eqref{resonance} with $ n=m \pm 1$, in which case it is given by ${\cal G} =\begin{pmatrix} 0 & \mp \widetilde{G}^j_{i}(k)/\left[4\mathrm{i}\omega_i(k)/\omega_d\right] \\ \widetilde{G}^i_{j}(k)/\left[4\mathrm{i}\omega_{j}(k)/\omega_d\right] & 0 \end{pmatrix}$. Thus, to first order, the perturbed eigenvalues are given by $s^{im\pm }+ \varepsilon_{ij}$ and $s^{jn\mp} - \varepsilon_{ij}$ with
\begin{equation}
\label{instability}
\varepsilon_{ij} = \frac{A_d}{4\omega_d^2} \sqrt{\pm\frac{\widetilde{G}^j_{i}(k)\widetilde{G}^i_{j}(k)}{\omega_i(k)\omega_j(k)/\omega_d^2}},
\end{equation}
where the $\pm$ in Eq.~\eqref{instability} is determined by the $\pm$ on the left hand side of Eq.~\eqref{resonance}.

The sign of $\pm\widetilde{G}^j_{i}(k)\widetilde{G}^i_{j}(k)$ determines whether the eigenvalues acquire a real component and thus result in an instability. When small damping is present, arbitrarily small $A_d$ will not result in eigenvalues with positive real part even if $\pm\widetilde{G}^j_{i}(k)\widetilde{G}^i_{j}(k)>0$, since the unperturbed eigenvalues then have an additional, strictly negative, real damping term. In this case, the driving amplitude must be sufficiently large for the perturbation in Eq.~\eqref{instability} to overcome the damping and result in an instability. Such instabilities may never occur if the damping is too large, as higher order terms in the perturbative analysis become relevant for large $A_d$.

For instabilities in Eq.~\eqref{instability} with $i=j$, Eq.~\eqref{resonance} implies $\omega_d=\omega_i(k)/2$, which corresponds to a subharmonic response. Harmonic responses appear only at second order in the degenerate perturbation theory and are therefore less prominent than subharmonic responses when small damping is present. Instabilities with $i\neq j$ result in an anharmonic response at first order with a frequency ratio $\epsilon = \omega_i(k)/\left[\omega_i(k)+\omega_j(k)\right]$ for $\omega_i(k)<\omega_j(k)$. However, if $2\omega_{i'}(k') = \omega_i(k) + \omega_j(k)$ for some $i'$ and $k'$, a subharmonic instability will occur at the same driving frequency as the anharmonic instability. If $4\widetilde{G}_{i'}^{i'}(k')^2\omega_i(k)\omega_j(k) >\pm{\widetilde{G}^j_{i}(k)\widetilde{G}^i_{j}(k)}\omega_d^2$, then the anharmonic instability will not be observable, since the subharmonic mode will go unstable at lower driving amplitude than the anharmonic modes. In the case that $\omega_d=\omega_i(k) + \omega_j(k)$ is twice the frequency corresponding to a band gap in the dispersion relation, there are no possible subharmonic responses, and the anharmonic response will generally be observed.

As an example, we now apply the degenerate perturbation theory to the pendulum array in Eq.~\eqref{springs}. In this case, there are two coupled fields corresponding to long and short pendula, and the undriven modes $\psi_0$ and $\psi_1$ correspond to acoustic and optical modes described in Fig.~\ref{fig1}(b). Figure \ref{fig7}(a) shows the branches of resonance and coresonance resulting in prominent instabilities for the pendulum array, as determined by the first order theory in Eq.~\eqref{resonance} with $m=n\pm1$. As shown in Fig.~{\ref{fig7}}(b), the drive-induced coupling matrix satisfies $\widetilde{G}_0^1(k)\widetilde{G}_1^0(k)>0$, and thus instabilities in Eq.~\eqref{instability} only occur for the coresonance condition in Eq.~\eqref{resonance} with the plus sign on the left hand side. The small damping in Eq.~\eqref{springs} perturbs the Floquet exponents $s^{im\pm}$ with a real component $-{\eta}/{2\omega_d}$. When the real part of the perturbation in Eq.~\eqref{instability} cancels this damping component, instability occurs, which specifies the first-order approximation of the instability boundary. The first-order theory approximates the instability boundary very well for resonant driving frequencies corresponding to the most prominent instabilities, as shown in Fig.~{\ref{fig7}}(c). Outside the frequency bands in Fig.~{\ref{fig7}}(a), the driving amplitude must be sufficiently large to produce degeneracy in the Floquet exponents, as we showed in Fig.~\ref{fig4}(a) using the non-perturbative solutions to Eq.~\eqref{eigen1}. Higher orders in perturbation theory are necessary to determine the stability boundaries for such driving frequencies, but anharmonic responses can generally occur at higher order as well, as was the case for the anharmonic tongue in Fig.~\ref{fig3}(b).

It is interesting to note that the nature of the alternating heterogeneity in the pendulum array can qualitatively change the instabilities by altering the drive-induced coupling $\widetilde{G}_i^j(k)$. For example, if the masses of the pendula alternate instead of the lengths, the coefficients of $\ddot{\phi}_i$ and $A_d \cos(\omega_d t)\phi_i$ in the equations of motion corresponding to Eqs.~\eqref{springs2} and \eqref{springs3} become identical, so that the drive-induced coupling matrix elements $G_i^j(k)$ become proportional to $\delta_i^j$. This results in $\widetilde{G}_0^1(k) \widetilde{G}_1^0(k) = 0$ for $i\neq j$, which implies that no anharmonic instabilities would occur. Similarly, if $G_i^j=F_i^j$ in Eq.~\eqref{coupled}, then it follows that $\widetilde{G}_i^j(k)\widetilde{G}_j^i(k)=0$ for $i\neq j$, so that anharmonic instabilities will not occur if the undriven coupling and the drive-induced coupling are identical. On the other hand, different forms of heterogeneity or parametric driving could, in principle, result in $\widetilde{G}_0^1(k)\widetilde{G}_1^0(k) < 0$, which would imply that anharmonic instabilities occur at the differences rather than the sums of natural frequencies in Eq.~\eqref{resonance}.

\section{Discussion}
\label{discussion}
Our demonstration of anharmonic instabilities in driven many-body systems reveals a fascinating form of classical discrete time crystals that more fully break the temporal symmetry of the periodic driving than the previously considered subharmonic instabilities. It is especially surprising that the anharmonic response can maintain its quasiperiodic coherence after nonlinear saturation, given that the relative phase between the driving force and the response varies from cycle to cycle for the typically irrational frequency ratios predicted. Floquet analysis and numerical simulations of the anharmonic response show that the resulting phase is indeed a rigid, collective phenomenon that emerges from a coresonance between the acoustic and optical modes in the array.

Anharmonic responses can occur in general in media with gapped dispersion relations that satisfy certain conditions. As we have shown, they can be characterized through a Neimark-Sacker bifurcation in the stroboscopic map{, which is} strobed at the driving frequency. Unlike the well-known torus bifurcation, however, they represent a transition directly between a fixed point and an invariant torus in the continuous-time dynamics, with no limit cycle acting as an intermediate between the two. Thus, anharmonic responses lie outside of the classical Cross-Hohenberg classification of pattern-forming systems characterized by a single critical wavelength and frequency \cite{1993_Cross_Hohenberg} and require new methods for analysis. We showed that the following are sufficient conditions for the emergence of anharmonic responses in the general: (i) the drive-induced coupling given by $\pm\widetilde{G}_i^j(k)\widetilde{G}_j^i(k)$ in Eq.~\eqref{instability} between different modes (which are decoupled in the absence of driving) must be sufficiently strong in order to overcome the damping; and (ii) the anharmonic instability must occur at lower driving amplitudes than any other resonant subharmonic modes for the given driving frequency, which can occur generally for driving frequencies corresponding to twice a frequency within a band gap.

In addition to banded systems arising from periodic compositions, such as the pendulum array and electronic states in crystals \cite{1972_Ziman}, other examples may include coupled electromagnetic and acoustic waves in plasmas \cite{2013_Kyoji}, piezoelectrics \cite{1949_Kyame}, and paraelectrics \cite{2019_Grimalsky}, as well as coupled flows in reaction-diffusion systems and interfacial fluid dynamics \cite{1993_Cross_Hohenberg, 2005_Pototksy}. We anticipate that anharmonic responses may be experimentally realized in simple systems with appropriate design enabled by our perturbative theory. For example, a vessel filled with water and oil will form a bilayer that can be modeled by coupled fields corresponding to the thickness of each fluid layer. Parametric driving by vertical vibrations will then result in an additional drive-induced coupling between the fields, which could result in anharmonic Faraday wave instabilities. We expect that the viscous theory for a single fluid interface \cite{1994_kumar} can be generalized to the bilayer case, and that Eqs.~\eqref{resonance} and \eqref{instability} will inform parameter design to create anharmonic Faraday wave instabilities in such bilayer systems.

We suggest that anharmonic responses can be useful in applications beyond the study of new phases of matter and new mechanisms of pattern formation. In particular, technologies that achieve tunable analog frequency conversion may be realized through anharmonic responses. Such technologies may find use in power-grid networks, where long-distance transmission at reduced frequencies can substantially reduce losses~\cite{2015_Liu}, and in mechanical metamaterials, where frequency responses can be manipulated for {applications such as} acoustic cloaking~\cite{2016_Ma,2019_Ronellenfitsch}.\smallskip

All essential data and code used in our simulations are available at the GitHub repository \cite{github}.

\begin{acknowledgments}
The authors thank Daniel J. Case and Chao Duan for insightful discussions. This work was supported by U.S. Army Research Office Grants No.\ W911NF-20-1-0173 and No.\ W911NF-19-1-0383 and Northwestern University’s Finite Earth Initiative (funded by Leslie and Mac McQuown).\\[0.0em]
\end{acknowledgments}

\begin{appendix}
\section{Numerical Integration}
\label{numericsec}
Integration is carried out in Python with $N$ pendula and periodic boundary conditions. Instabilities are then limited to wave numbers $k=n\pi/N$ for integers $0\leq n\leq N-1$. Instabilities saturate to a steady response after a few hundred oscillation periods for driving amplitudes just above the instability boundary determined by Floquet analysis. The frequency components of the saturated instability agree precisely with those given by the Floquet exponents $\omega_d\epsilon$ and $\omega_d(1-\epsilon)$. We approximate the Gaussian white noise by sampling from a Gaussian distribution every $\tau$ time units with a variance $\sigma^2/\tau$, which produces a piecewise noise tending to white noise in the limit of small $\tau$. 

\end{appendix}

\end{document}